\documentstyle{article}
\newcommand{\Be}{\begin{equation}}
\newcommand{\Ee}{\end{equation}}
\newcommand{\ybar}{\bar{\psi}}
\newcommand{\n}{\nu}
\newcommand{\y}{\psi}
\newcommand{\half}{{\scriptsize\frac{1}{2}}}
\newcommand{\dslash}{/\!\!\!\partial}
\newcommand{\Bea}{\begin{eqnarray}}
\newcommand{\Eea}{\end{eqnarray}}
\newcommand{\Eq}[1]{equation~(\ref{#1})}
\newcommand{\Mn}{m^{(0)}_\n}
\newcommand{\Ra}{\rightarrow}
\newcommand{\Ealf}{\nonumber\\}
\begin{document}
\thispagestyle{empty}
\centerline{\normalsize\bf }
\baselineskip=22pt
\centerline{\normalsize\bf Neutrino Clouds}
\baselineskip=13pt
\centerline{\footnotesize G. J. Stephenson Jr.}
\baselineskip=13pt
\centerline{\footnotesize\it Department of Physics \& Astronomy,
 University of New Mexico}
\centerline{\footnotesize\it Albuquerque, New Mexico 87131}
\vspace*{0.3cm}
\centerline{\footnotesize and}
\vspace*{0.3cm}
\centerline{\footnotesize T. Goldman}
\baselineskip=13pt
\centerline{\footnotesize\it  Theoretical Division, Los Alamos  
National
Laboratory}
\baselineskip=12pt
\centerline{\footnotesize\it Los Alamos, New Mexico 87545}
\vspace*{0.3cm}
\centerline{\footnotesize and}
\vspace*{0.3cm}
\centerline{\footnotesize B. H. J. McKellar}
\baselineskip=13pt
\centerline{\footnotesize\it School of Physics, University of Melbourne}
\baselineskip=12pt
\centerline{\footnotesize\it Parkville, Victoria 3052, Australia}

\begin{abstract}
We consider the possibility that neutrinos are coupled very weakly to
an extremely light scalar boson.  We first analyze the simple problem
of one generation of neutrino and show that, for ranges of parameters
that are allowed by existing data, such a system can have serious
consequences for the evolution of stars and could impact precision
laboratory measurements.  We discuss the extension to more generations
and show that the general conclusion remains viable.  Finally, we note
that, should such a scalar field be present, experiments give
information about effective masses, not the masses that arise in
unified field theories.
\end{abstract}

\pagebreak
\setcounter{page}{1}

\section{Introduction}
In the last few years, it has been suggested that neutrinos might
interact weakly among themselves through the exchange of a very light
scalar particle~\cite{MKY,TRE}, with possible consequences for the
evolution of the Universe and for the propagation of neutrinos from
distant events.  In many of these discussions, one assumes that
neutrinos are distributed according to the usual Big Bang scenario and
Standard Model physics, the effects of scalar exchange being treated as
a perturbation.  In this paper we examine that assumption.

This problem is a special case of the general problem of relativistic
fermions interacting through the exchange of scalar and vector bosons,
and the general formalism has been worked out under the name Quantum
Hadrodynamics (QHD) and applied to the study of nuclear
physics~\cite{QHD}.  We shall show that, for a wide range of
parameters, neutrinos will tend to cluster, that these clusters could
be of a size to affect stellar formation and dynamics and that there
may be observable consequences for physics within the solar system.  In
fact, if the clustering is strong enough so that the density of
neutrinos within a cloud is sufficiently large, terrestrial laboratory
experiments can be affected.

Over the last decade, many groups studying the endpoint of the Tritium
beta ray spectrum for signs of neutrino mass have reported a best fit
value for the square of the anti-neutrino mass less than
zero~\cite{TRIT1,TRIT2,TRIT3,TRIT4,TRIT5,TRIT6}.  Robertson
et.al~\cite{TRIT1} point out that this result could be obtained by
assuming a very small branch due to the absorption of relic neutrinos,
provided the density of such neutrinos were some $10^{13}$ higher than
usual cosmological values.  In choosing parameter ranges for the
examples in this paper, we have kept this idea in mind.  We must
emphasize, however, that the general feature of cloud formation and its
consequence for the evolution of structure in the history of the
Universe is quite robust, and must be considered whatever the eventual
resolution of the anomaly in Tritium beta decay.

In this work, we only consider the effects of light scalar exchange.
The effects of the known vector exchange ($Z_0$) are far too small
(really, too short ranged) to affect these results.  The exchange of a
light vector particle is severely constrained by data.  To avoid
problems with the axial anomaly (the neutrinos do couple to the $Z_0$)
one must either invent many new fermions or demand that the light boson
couple to known leptons or quarks, which quickly leads to conflict with
experiments designed to test for a fifth force~\cite{EA}.  Furthermore,
the self shielding of a vector exchange, while it might allow for the
development of a neutrino-antineutrino plasma, will not allow for the
coherent action required to drive cloud formation.  Thus, we only treat
scalar exchange.  Even so, there remains the rich possibilities of
different couplings to different generations. Aside from a few
comments, we leave that to further work, concentrating our discussion
on the simpler system of one flavor of neutrino.

The paper is organized as follows.  In section 2 we review the
treatment of infinite matter in QHD and apply that to the problem at
hand.  In section 3 we do the same for finite clouds of neutrinos.  In
section 4, we describe the consequences such clouds would have on the
the evolution of structures in the Universe and, in section 5, we
confront this picture with what data exists, extracting limits on the
parameters.  In section 6 we discuss the changes in the analysis of
Tritium beta decay experiments within such neutrino clouds, and comment
on the effects of such clouds on stellar dynamics in section 7.  In
section 8 we discuss some aspects of the extension to include more than
one generation, illustrating the remarks with special cases applied to
two generations.  We offer our conclusions in section 9.

\section{The Infinite Problem}

The Lagrangian for a Dirac Field interacting with a scalar field is
well known:
\Be 
{\cal L} = \ybar(i\dslash - m_\n)\y +\half\left[\phi(\partial^2 -
m_s^2)\phi\right] +g\ybar\y\phi 
\Ee
which gives as the equations of motion
\Bea
\left[\partial^2 + m_s^2\right]\phi & = & g\ybar\y \label{PHI}\\
\left[i\dslash - m_\n\right]\y & = & -g \phi \y.\label{PSI}
\Eea
We omit nonlinear scalar selfcouplings here, even though they are
required to exist by field theoretic selfconsistency, as they may
consistently be assumed to be sufficiently weak as to be totally
irrelevant.

We look for solutions of these equations in infinite matter which are
static and translationally invariant. Equation (\ref{PHI}) then gives
\Be
\phi = \frac{g}{m_s^2} \ybar\y,
\Ee
which, when substituted in (\ref{PSI}) gives an effective mass for the
neutrino of
\Be
m^*_\n = m^{(0)}_\n - \frac{g^2}{m_s^2}\ybar\y.
\Ee
where $m^{(0)}_\n$ is the renormalized vacuum mass that the neutrino
would have in the absence of other physical neutrinos.

These equations are simply the equations of Quantum Hadrodynamics
\cite{QHD}, and we will be using them in a small coupling regime where
there is no question of the validity of neglecting higher order
processes.

These equations are operator equations. We next act with each of these
equations on a state $|\Omega \rangle$ defined as a filled Fermi sea of
neutrinos, with a number density $\rho$, and Fermi momentum $k_F$,
related as usual by $\rho~=~{k_{F}}^{3}/(6\pi^{2})$.  The operator
$\ybar\y$ acting on this state gives
\Be
\ybar\y |\Omega \rangle = \frac{w}{(2\pi)^3}\int_{|\vec{k}|< k_F}  
d^3k
\frac{m^*_\n}{\sqrt{(m^*_\n)^2 + k^2}} |\Omega \rangle,
\Ee
where $w$ is the number of neutrino states which contribute --- $w=2$
for Majorana neutrinos and $w=4$ for Dirac neutrinos.   Thus the
effective mass is determined from the integral equation
\Be
m^*_\n = m^{(0)}_\n - \frac{g^2w}{2\pi^2m_s^2}
\int_0^{k_F} k^2\,dk
\frac{m^*_\n}{\sqrt{(m^*_\n)^2 + k^2}.}\label{MSTAR}
\Ee

To discuss the solutions of this equation we reduce it to dimensionless
form, dividing by $m^{(0)}_\n$, and introducing the parameter
$K_0~=~\frac{g^2w(m^{(0)}_\n)^2}{2\pi^2m_s^2}$, and the variables
$y~=~\frac{m^*_\n}{m^{(0)}_\n}, x~=~\frac{k}{m^{(0)}_\n},
x_F~=~\frac{k_F}{m^{(0)}_\n}$.  Then \Eq{MSTAR}  becomes
\Bea
y& = &1 - y K_0 \int_0^{x_F} \frac{x^2\,dx}{\sqrt{y^2+x^2}}\\
&=& 1 - \frac{y K_0}{2}\left[e_Fx_F - y^2 \ln\left(\frac{e_F +
x_F}{y}\right) \right],\label{YX}
\Eea
with $e_F~=~\sqrt{x_F^2 + y^2}$.  One can regard \Eq{YX} as a
non-linear equation for $y$ as a function of either $e_F$ or $x_F$.  As
a function of $e_F$, $y$ is multiple valued (when a solution exists at
all), whereas $y$ is a single valued function of $x_F$.

The total energy of the system is a sum of the energy of the neutrinos,
$ E_\n~=~e_\n~\Mn~wN$, and the energy in the scalar field,
$E_s~=~e_s~\Mn~wN$, where $N$ is the total number of neutrinos in each
contributing state.  These expressions serve to define the per neutrino
quantities $e_\nu$ and $e_s$.  Also, $E_s~=~{\cal E}_s~V$, where ${\cal
E}_s~=~\frac{1}{2}m_s^2\phi^2$ is the energy density of the (here
uniform) scalar field.

One finds that
\Bea
e_\n & = & \frac{3}{x_F^3}\int_{0}^{x_F}x^2\, dx \sqrt{x^2 +  
y^2}\Ealf
& = & \frac{3}{x_F^3}\left\{\frac{x_F^3e_F}{4} + \frac{x_Fy^2  
e_F}{8} -
\frac{y^4}{8}\ln\left(\frac{e_F + x_F}{y}\right)\right\}
\Eea
and
\Bea
e_s & = &  
\frac{K_0}{2}\frac{3}{x_F^3}y^2\left(\int_{0}^{x_F}\frac{x^2\,
dx}{\sqrt{x^2 + y^2}}\right) ^2\Ealf
& = & \frac{1}{2 K_0} \frac{3}{x_F^3} (1-y)^2.
\Eea

Notice that for large values of $x_F$, 
\Bea
y & \Ra & \frac{2}{K_0 x_F^2}\Ealf
e_\n & \Ra & \frac{3 e_F}{4}\Ealf
& \Ra & \frac{3 x_F}{4}\Ealf
e_s & \Ra & \frac{3}{2K_0} \frac{1}{x_F^3}.
\Eea

It is also useful to note that, for small $x_F$, 
\Bea
y & \Ra & 1 - \frac{K_0 x_F^3}{3}\Ealf
e_\n  & \Ra &  1 + \frac{3 x_F^2}{10}\Ealf
e_s  & \Ra &  \frac{K_0}{2}\frac{x_F^3}{3}.\label{LEL}
\Eea

For the neutrino system to be bound, the minimum of $e~=~e_\n~+~e_s$ as
a function of density (or $x_F$) must be less than 1, its value in the
zero density limit.

These points are illustrated in Figures 1 and 2, for the case
$K_0~=~20$.  In Figure 1, the average energy per neutrino, $e$, and the
Fermi energy, $e_F$, are plotted, in units of the vacuum mass $m_0$,
against the Fermi momentum in the same units, $x_F$.  In Figure 2, the
same information is displayed as $e$ vs $e_F$, demonstrating the
multivaluedness of the solution viewed this way.  The general shape of
Figure 1 will be exhibited for any value of $K_0$, but only if $K_0$ is
large enough will there be a minimum with $e < 1$.  This is displayed
in Figure 3, from which we may deduce that $K_0$ must be greater than
$\approx 3.3$ for binding to occur.  Note also that the local minimum
disappears for $K_0~=~2.67$, showing that there is no metastable state
for smaller values of the effective coupling strength.

\section{Neutrino Clouds}

Note from \Eq{LEL} that the energy varies as $1 + \mbox{const.}\times
k_F^2$ near $x_F~=~0$, where $const.$ is positive.  Even with $K_0$
large enough to produce binding, at low density $e$ will be greater
than $1$, then will decrease with increasing $x_F$ to its minimum value
before increasing as $x_F~\rightarrow~\infty$.  Thus, for low
densities, it may be energetically favourable for the neutrinos to form
finite size clouds.

To investigate this possibility we use the Thomas-Fermi approximation
to obtain the scalar field through the equation
\Be
\nabla^2 \phi + m_s^2\phi = g \langle \ybar \y \rangle, \label{LDE}
\Ee
substituting the local density value for $\langle \ybar \y \rangle$.

Noting that
\Be
y = 1 -\frac{g}{\Mn} \phi,
\Ee
we convert the local density  \Eq{LDE} for $\phi$ to a local density
equation for $y$:
\Be
\frac{d^2y}{dz^2} + \frac{2}{z} \frac{dy}{dz} =  
G(y,e_F,K_0)\label{CDE}
\Ee
with 
\Be
G(y,e_F, K_0) = -1 + y\left(1+K_0\int_0^{x_F} \frac{x^2\,  
dx}{\sqrt{x^2 +
y^2}}\right),
\Ee
and $z~=~r~m_s$, where $r$ is the radial distance from the center of
the cloud and we have assumed spherical symmetry.  In this equation,
$e_F$ (which is a chemical potential for the neutrinos) is constant
with position, but $y$ and thus $x_F~=~\sqrt{e_F^2~-~y^2}$ are
functions of position.  Clearly the surface of the cloud is defined by
the condition that $x_F~=~0$, as there the neutrino density drops to
zero, at the (scaled) radius $z~=~z_0$.  For $z~>~z_0$, $x_F$ remains
zero, although the scalar field does not immediately vanish.  Instead,
outside the cloud the solution of the differential equation for the
effective mass $y$ becomes
\Be
y = 1 - (1-e_F) \frac{e^{-(z-z_0)}}{z/z_0}\label{EXT}
\Ee
which satisfies the condition  $y(z_0)=e_F$ at the surface, and the
condition that $y \Ra 1$ as $z \Ra \infty$.  From this exterior
solution we compute the values of $y$ and $\frac{dy}{dz}$ at $z=z_0$,
to which the interior solution must be matched:
\Bea
y(z_0) & = & e_F\Ealf
\left.\frac{dy}{dz}\right|_{z=z_0}& =&
\left(1-e_F\right)\left(1+\frac{1}{z_0}\right).
\Eea

In the usual way one solves the interior equation numerically starting
from an assumed central value, $y(0)$, which is adjusted to satisfy the
matching conditions.  While this technique works in a straightforward
way for small clouds, as the cloud increases in size one needs to know
$y(0)$ to greater and greater precision, and needs to adopt a special
technique to handle such cases numerically.

For large values of $z_0$ the central region will approximate the
conditions of the infinite system, and so we expand the  function $G$
of \Eq{CDE} as a Taylor series in $y$ about $y_0$, the solution of
$G(y,e_F,K_0)~=~0$, which is the effective mass in the infinite
system.  Explicitly
\Bea
G(y,e_F,K_0) &=& \left\{-2 +\frac{3}{y_0} -
K_0e_Fx_0\right\}\left(y-y_0\right) \Ealf
&+& \left\{\frac{3}{y_0^2} -
\frac{3}{y_0} + \frac{K_0e_F}{2}\left(\frac{y_0}{x_0} -
3\frac{x_0}{y_0}\right)\right\}\left(y-y_0\right)^2 + \ldots.
\label{TAYLOR} 
\Eea

Keeping just the first term in \Eq{TAYLOR}, the \Eq{CDE} has the
solution
\Be
y - y_0 = A\frac{\sinh(\kappa z)}{\kappa z},\label{INT}
\Ee
where
\Be
\kappa^2 = \left\{-2 +\frac{3}{y_0} - K_0e_Fx_0\right\}.
\Ee
(It is straightforward to show that $\kappa^2 > 0$ for the cases of
interest, which are those for which $e$ is close to the lower branch in
Figure 2.)

The numerical solution of \Eq{CDE} is started at $z~=~z_1~>~0$, with an
assumed value $y(z_1)~=~y_0~+~\Delta y$, where $\Delta y$ is small
enough that the higher order terms in \Eq{TAYLOR} can be safely
neglected.  The value of $z_1$ is adjusted to satisfy the outer
matching conditions.  One can then use \Eq{INT} to extrapolate $y(z)$
from $z~=~z_1$ to $z~=~0$, obtaining
\Be
y(0) = y_0 + \frac{\kappa z_1}{\sinh(\kappa z_1)} \Delta y.
\Ee
In this way we can set $y(0)$ at the extraordinary level of precision
required to satisfy the matching conditions.

The general form of the solutions does not depend on the particular
value of $K_0$.  An example of solutions with different numbers of
neutrinos is presented in Figure 4 for $K_0~=~20$. To construct the
Figure, values of $e_F$ were chosen and a solution obtained, giving the
density as a function of the radial distance from the center of the
cloud.  That density was then integrated to obtain the total number N
displayed with the appropriate curves. The units are such that N=1
would correspond to $(m_\nu^0/m_s)^3$ total neutrinos.  For the
Thomas-Fermi approximation to work well, that number should be large
compared to 1, which, as will be seen below, is the case.  For later
reference, we display the radial form of the gravitational force,
$f_g$, due to such a distribution in Figure 5. This force is presented
in arbitrary units, since the masses have been scaled out, and is
calculated as $f_g~~=~\frac{1}{z^2}~\int_{0}^{z}~e~z^2~dz$.  Since, for
finite clouds, the scalar field is no longer always constant, there is
an additional contribution\cite{QHD} to the scalar field energy
density, $\frac{1}{2} ( \nabla \phi )^2$, which has been included in
our calculation.

To understand the distribution of cloud sizes to be expected, one needs
to know the average energy per neutrino as a function of size or of
total number.  For $K_0~=~200$, Figure 6 displays both the logarithm of
the cloud radius (i.e. the radius at which the density goes to zero)
and the logarithm of the difference between the average energy in the
cloud and the average energy in infinite neutrino matter at its minimum
for the same value of  $K_0$ versus the logarithm of the total number.
The points are the results of calculations corresponding to the
indicated values of $e_F$.  The line through the radius values is a
straight line with slope $\frac{1}{3}$, that through the energy
differences has slope  $\frac{-1}{3}$.  Evidently, for large N, the
radius scales as $\mbox{N}^{1/3}$ and the energy difference scales as
$\mbox{N}^{-1/3}$, appropriate to a surface tension, indicating a
preference for large clouds.  The line through the $e_F$ values is to
guide the eye.

\section{Early Universe}

Consider the effects of such clustering on the evolution of structures
in the early Universe.  Throughout the following discussion, we assume
that $K_0$ is large enough to produce bound systems.  At an early
enough epoch the density will be sufficiently high that the effective
mass is negligible.  At that epoch, there is no difference between the
interacting neutrinos and the relativistic, non-interacting gas that is
usually assumed.  Consequently, these neutrinos will expand and
decrease in density according to the standard scenario until the
increase in the effective mass begins to make a difference, which will
occur at about the value of the density ($x_F$) corresponding to the
minimum energy per particle for infinite matter.

Were it possible to remove energy from the neutrino gas and entirely
from the Universe, so that the gas could be viewed as having zero
temperature, that would be the end of the discussion.  We would,
however, be left with a conundrum.  The neutrinos could tolerate no
further expansion but the Universe, being driven by all sources of
energy density, would be required to continue to expand. This would
result in one neutrino cloud located somewhere (defining a ``center''
?).

That, of course, is not the situation.  The neutrinos will have a
temperature comparable to that obtained for an expanding,
non-interacting gas.  As the expansion continues, that temperature will
be converted into (effective) mass and the gas will become supercooled,
followed by fragmentation into clouds.  Note that no additional
dissipation is required, unlike the case where clouds coagulate from
free particles.  The point here is that the neutrinos were born within
a cloud and never achieve a state in which the effective mass rises to
its vacuum value.

Many neutrinos have been born at later times through normal stellar
burning, supernova explosions or other processes.  When they encounter
a neutrino cloud, the coherent forward scattering is easily large
enough, even for very small values of the coupling to the scalar field
so that individual scatterings are small, to cause the neutrinos to
lose energy through the Bremsstrahlung of scalars, providing additional
dissipation.

Two factors drive the size distribution of these clouds.  The first is
the distribution of fluctuations, which we assume follows
Harrison-Zel'dovich~\cite{HZ}.  The second is the increase in energy
per neutrino with decreasing cloud size discussed in section 3.  The
latter effect provides for a mechanism to cut off the distribution of
cloud sizes below some smallest value, the actual efficacy of which
depends on the detailed parameter values.  The general form of the
distribution is
\begin{displaymath}
P(N) \propto N^{-2}\exp(-C/N^{1/3}).
\end{displaymath}

Should this process occur before recombination, which, as we shall see
in Section 5, is not beyond reason, then the existence of neutrino
clouds would have a profound effect on the evolution of small size
structures.  (By small, in this context, we refer to structures of the
size of solar systems, stars or a bit smaller.)  At recombination, when
matter decouples from the background photon gas, there will be a
pre-existing collection of gravitational sources.  The longer the time
between cloud formation and recombination, the more these will appear
to be point sources, but that does not strongly affect the following
argument.  Whatever the spectrum of fluctuations in the baryon
distribution, these pre-existing sources will nucleate baryon
condensation with a distribution that more or less follows the size
distribution of the clouds.

Many of these collections of baryons will be large enough to initiate
nuclear burning and become stars; others will not.  Of the latter, some
will attract more baryonic matter from the ejecta of exploding stars to
form later generation stars, while others will remain too small to
evolve into stars and can provide cold, massive objects.  Note that,
even if a given cloud does not attract a compliment of baryonic matter,
it will still function as a gravitational source.  In either case, the
increase in the energy per neutrino with decreasing cloud radius,
discussed above, will provide a lower limit to the distribution of
system sizes.  Thus, the existence of neutrino clouds can serve as a
seed mechanism for stars and could provide a similar seed for (or be
themselves) smaller objects such as MACHOs\cite{AA}.

This scenario suggests that all stars will have their associated
neutrino cloud, not because stars attract neutrinos but, rather,
because stars form within the gravitational well provided by
pre-existing neutrino clouds.  One may then ask if, during subsequent
evolution, the star and its cloud remain together or if the star,
buffeted by forces which ignore the neutrinos, is stripped away leaving
the cloud to catalyze another object.  This is a quantitative question
which we discuss in section 7.

\section{Numerical Considerations}

Following the spirit outlined in the introduction, we shall consider
here the constraints that may be placed on the parameters of the theory
under the assumption of one surviving neutrino species.  We shall take
up the question of other generations in section 8.

First, to impact terrestrial Tritium beta decay experiments, the
density of electron neutrinos needs to be about $10^{15}/cm^{3}$.
Robertson et.\ al.\ \cite{TRIT1} reported that their data could be fit
by assuming an additional branch, $10^{-9}$ of the normal decay, which
fed each final state in the same proportion as the normal decay.
Applying standard formulas~\cite{AFH} for inverse beta decay, this rate
requires a density of $6\times 10^{15}/cm^{3}$.  A recent paper by the
Moscow group~\cite{TRIT6} indicates a preference for a $6\times
10^{-11}$ branch, which would imply a density of
$3.6\times~10^{14}/cm^{3}$.

For the first case,
\begin{displaymath}
k_F = (6\pi^2\rho)^{1/3}
= 14 eV/c
\end{displaymath}
while the second would imply
\begin{displaymath}
k_F=5.5 eV/c
\end{displaymath}
Throughout this section we shall use 10 eV/c, since we are arguing here
for the scale of the parameters.  (An actual fit requires a more
complete treatment of the modification of the spectrum, which we
discuss further in section 6).  This value of $k_F$ corresponds to a
density of $2\times 10^{15}/cm^{3}$, which is some thirteen orders of
magnitude above that obtained in standard cosmology, assuming that
neutrinos are uniformly distributed in space.  If we assume that all
neutrinos created when the weak interaction was sufficiently strong to
keep thermal contact between neutrinos and photons survive to the
present epoch, this implies that only $10^{-13}$ of space is occupied
by neutrinos.

As a consequence, neutrinos propagating over galactic or inter-galactic
distances will, in the main, be free of the influence of condensed
neutrinos and will propagate with their vacuum mass.  Therefore, mass
limits obtained from the spread of arrival times for neutrinos from
SN1987a will apply to the vacuum mass~\cite{SN87a}. While lower limits
have been obtained by some authors, we shall use a limit of 50 eV
here.  Coupled with $k_F~=~10 eV$, this implies
\begin{displaymath}
x_F>0.2
\end{displaymath}
or
\begin{displaymath}
K_0<4000 ~~.
\end{displaymath}

From the requirement that the system be bound, we obtain
the lower limit
\begin{displaymath}
K_0>3.3 ~~. 
\end{displaymath}
For one generation, particle and antiparticle (chiral left and right,
if Majorana), $K_0$ is related to the basic parameters by
\begin{equation}
K_0=4\tilde \alpha/\pi\mu^2
\end{equation} 
where
\begin{equation}
\mu = m_s/m^{(0)}_\nu
\end{equation}
and
\begin{equation}
\tilde \alpha = g^2/4\pi ~~.
\end{equation}

To obtain limits on $\tilde \alpha$ directly, we consider those
processes involving either the scattering of neutrinos through scalar
exchange or processes involving the Bremsstrahlung of scalars.  For the
purpose of these estimates, we use only leading terms, and it does not
matter whether the neutrino is a Dirac or Majorana particle.
Consequently, many existing limits on the coupling of a Majoron to
neutrinos can be taken over directly.

In a study of the limits on such a coupling in a number of elementary
particle processes~\cite{GKS}, it was found that the most stringent
limit was that given by Barger, Leung and Pakvasa~\cite{BLP} from a
study of $(K_{l2})$ decay.  This gave 
\begin{displaymath} 
\tilde \alpha < 10^{-6} ~~.  
\end{displaymath} 
A more stringent limit can be obtained from recent studies of nuclear
double beta decay~\cite{DBL1} looking for the spectral signature which
would accompany the emission of one extra particle.
\begin{displaymath} \tilde \alpha < 10^{-9} ~~.  \end{displaymath} Note
that this limit only applies if neutrinos are Majorana particles.

For $\surd s \gg m_\nu$ the total neutrino-neutrino scattering
cross-section is given by
\begin{displaymath}
\sigma \approx 5\pi\tilde \alpha ^2/s ~~.
\end{displaymath}
The strongest constraint from scattering comes from the neutrinos
produced by SN1987a.  If we assume that all neutrinos born in the Big
Bang survive and are distributed roughly proportionally to the baryons,
then there are $10^{66}$ associated with the solar system.  At $2\times
10^{15}/cm^{3}$, that implies a radius of the cloud of $\approx 5\times
10^{16} \mbox{cm}.$  On this scale, the Earth is essentially at the
center of the cloud.  Thus the mean free path for 20 MeV neutrinos
coming in from the supernova must be greater than $5\times
10^{17}\mbox{cm}$ or the cross section must be less than
$10^{-34}\mbox{cm}^2$.  For a mean neutrino energy of 7.5 eV ($3k_F/4$)
and a beam neutrino of 20 MeV, $s \approx 1.5 \times 10^{-10}
\mbox{GeV}^2 $, giving $\sigma \approx10^{-17}\mbox{cm}^2\tilde
\alpha^2$  This then requires that
\begin{displaymath}
\tilde \alpha < 3\times 10^{-9} ~~.
\end{displaymath}

Other processes, such as the observation of solar neutrinos or the
observation of neutrinos from high energy accelerators lead to less
stringent upper bounds.

A more interesting bound comes from the consideration of the survival
of Big Bang neutrinos to the present epoch.  The dominant disappearance
mechanism is for a neutrino and an antineutrino to annihilate into a
pair of scalars.  For Majorana neutrinos, this simply means that a
chiral left and a chiral right neutrino transmute into a pair of
scalars.  The rate is
\begin{displaymath} 
\omega \approx  \frac{3}{8} \tilde\alpha ^2 k_F \ln \frac{k_F}{m_s}.
\end{displaymath}
The total disappearance probability is the integral of this rate to the
present from whatever initial time is appropriate.  The time before
neutrinos condense into clouds is irrelevant to this question as,
throughout that time, the scalar field and the neutrinos are both
expanding relativistic gases which were once in thermal contact.  All
that a large rate for this process can do is maintain that thermal
contact since, under these conditions, the back rate is the same.
Consequently, we consider the integral of the rate from a time when the
temperature corresponds to about 10 eV, or $t~=~10^{13} s$.  The bound
is only weakly dependent on the assumption of radiation or matter
dominance for the expansion.  Requiring a decrease of less than 0.1 in
$\ln(N/N_0)$ yields
\begin{displaymath}
\tilde \alpha < 3 \times 10^{-20}  \mbox{(radiation-dominated)}
\end{displaymath}
or
\begin{displaymath}
\tilde \alpha < 10^{-18}   \mbox{(matter-dominated)}.
\end{displaymath}

It is tempting to consider scenarios with $\tilde \alpha$ above this
limit but still below the limits imposed by scattering.  In that case,
one might argue that neutrinos are Dirac particles and that there is a
fractional particle excess as for any other fermion.  It is, in fact,
such a scenario that would give only an excess of counts at the end
points of beta decay spectra (see the discussion in the next section).
However, the limits on neutrino generations from nucleosynthesis
calculations are based on the assumption of only left handed neutrinos
and right handed anti-neutrinos.  The scalar interaction considered
here will equilibrate left and right handed neutrinos while they are in
equilibrium through the weak interaction, leading to a factor of two
too many states per generation.  As the remarks above regarding the
limits of the time integration for disappearance point out, it does not
seem possible to try to argue for the disappearance of some states at
an early enough epoch that it could affect the neutron-proton
equilibrium and, hence, nuclear abundances.

At this juncture, one may question one's ability to accommodate the
scalar field in light of the same argument.  This question was
addressed several years ago by Kolb, Turner and Walker~\cite{KTW}.  The
results of that paper indicate that a very light scalar (the case here)
would behave as an extra half a generation, which can be accommodated.

Returning to the issue of Majorana neutrinos, the search for the
evidence for their existence in nuclear double beta decay is a thriving
industry~\cite{DBL1,HXS}.  If such decays were mediated by a mass term,
modern experiments limit that mass to less than 1-2~eV.  When taken
with our standard value of $k_F$~=~10~eV, this implies that
$y/x_F~<~0.2$ independent of the vacuum mass.  Table 5.1 lists, for
some representative values of $K_0$, values of $x_F$, $y$ and the ratio
$y/x_F$ for infinite matter at its minimum energy per neutrino.
\begin{table}
\caption{Infinite Matter Values}
$$
\begin{array}{rrrrr}
K_0 & <e> & x_F & y & x_F/y \\
20 & .704 & .675 & .2 & .3 \\
200 & .410 & .405 & .06 & .15 \\
2000 & .233 & .232 & .0186 & .08
\end{array}
$$
\end{table}
This suggests a preference for larger values of $K_0$ within the range
discussed above.

Another possible source of constraint is the gravitational attraction
that such a neutrino cloud would create due to its energy density (note
that the static scalar field contributes here).  In section 3 the
gravitational acceleration due to various clouds was displayed.  Nieto,
et al~\cite{MMN} have recently discussed an anomalous acceleration
observed on the Pioneer spacecraft, essentially constant from 10 to 50
AU with a value of $10^{-9} m/s^2$.  While Figure 5 raises the
possibility of a nearly constant acceleration over a wide range of
distances, the magnitude would require an average energy, at a density
of $2\times 10^{15}/cm^{3}$, of $\approx 50 eV$, far in excess of the
values considered here.  For this discussion (one generation only),
this implies that no useful constraints are likely from gravity.  On
the other hand, such considerations add strength to the argument that
the range of the interaction ought to be of the order of several AU,
since the extent of the surface is given by that range.  Even though
the clouds could be much larger than the scalar range, such clouds
would have a relatively sharp surface and would not produce a radial
dependence that was gentle enough to appear constant.

Therefore, assuming the range of the interaction to be of the order of
1 AU give or take a few orders of magnitude, since
\begin{displaymath}
1AU \approx 1.5\times 10^{13} cm 
\end{displaymath} 
we have
\begin{displaymath} m_s(1AU) \approx 1.3\times 10^{-18} eV
\end{displaymath}
which suggests that we take
\begin{displaymath}
10^{-21} < \mu < 10^{-17} ~~.
\end{displaymath}
Thus, for a given $K_0$ , we have a range of allowed $\tilde \alpha$,
as may be deduced from Figure 7.  On that figure, the solid vertical
line represents the $\mu$ appropriate to a range of $1AU$ and
$m^{0}_\nu~=~10~eV$, the dotted lines on either side representing a
change of 2 orders of magnitude.  The diagonal lines are lines of
constant $K_0$, the right hand one for $K_0~=~4$ and the left hand one
for $K_0~=~40,000$.  The other constraints already discussed appear as
horizontal lines representing the various upper limits on
$\tilde\alpha$.  For $\mu$ and $K_0$ in the ranges above, we would
obtain
\begin{displaymath}
10^{-42} < \tilde \alpha < 10^{-31}
\end{displaymath}

Such small values of $\tilde \alpha$ trivially satisfy all other
operative constraints.

\section{Tritium beta decay}

As remarked in section 5, a scenario which  led to a (smaller) cloud of
Dirac neutrinos would provide for a more natural explanation of
additional counts at the end point of a beta decay spectrum than would
a scenario involving a larger number of Majorana neutrinos.  The
reason, which is well known~\cite{SWB}, is summarized here.

The endpoint of the spectrum, calculated from the energy release in the
decay, is not affected by the presence of the background neutrinos.
Electrons emitted following the absorption of a neutrino will have an
energy beyond the endpoint equal to the total energy of the neutrino.
Thus, for a cold gas, the spectrum will reflect the Fermi-Dirac
distribution extending from $E_0~+~m^*$ to $E_0~+~e_F$ where $E_0$ is
the true endpoint. (One may safely neglect the change in the field
energy due to the disappearance of one neutrino.)  If the neutrinos are
Majorana, there will be an equal distribution of anti-neutrinos which
will block the emission of the lowest energy anti-neutrinos from the
normal beta decay.  Since the distributions are equal, the blocking
will exactly balance the additional emission, leading to no net
increase in the rate.

In either case (Dirac or Majorana), the change in the spectrum is more
complex than the  simple addition of a spike.  The density required to
affect the experiments produces a Fermi momentum of at least several
electron volts, and the distribution of additional events must reflect
that.  It may well be that fitting with the correct spectral shape will
destroy the feature, reported in~\cite{TRIT1}, that a fit, equally
acceptable to that with a negative mass squared, can be achieved.  It
is also possible that the correct shape will not change or improve the
fit.  That question can only be resolved by fitting actual data with
all experimental effects included.  Note that the analysis described
here, as envisaged by~\cite{SWB}, is for one neutrino, assumed to be
the electron neutrino.  We return to this in section 8.

\section{Cloud Dynamics}

To analyze the dynamics of the system of cloud plus star in a galactic
environment, and, in particular, to determine if a star stays within
the cloud that seeded its formation, would require a modified
Fokker-Planck treatment~\cite{BINTRE}.  While a complete treatment
would require the complications of at least three generations of
neutrinos, we consider only one generation, to illustrate some of the
issues.

The primary mechanism for altering a star's trajectory is the
gravitational scattering between two stars that pass relatively near
each other and, to lowest order, the clouds simply follow along.  Since
the cloud-star system is polarizable, there will be an induced
dipole-dipole interaction, analogous to atomic scattering, which will
produce a Van der Waals like interaction.  While this may alter the
specifics of the velocity distribution slightly, it should not have a
major impact on the issue of the cloud remaining with its star.

A more serious question involves the interaction between two clouds
when they touch.  If we assume that the cloud contains an energy
equivalent to about $1M_\odot$, uniformly distributed to
$10^{17}~\mbox{cm}.$, then the gravitational binding energy of the star
to the cloud is $\approx~10^{39}~\mbox{eV}$ or about
$10^{-27}~\mbox{eV}/\nu$. According to Figure 6 for $K_0~=~200$, the
surface contribution to the energy per neutrino is
$\approx~10^{-3}m_0/\mbox{N}^{1/3}$.  Thus, for the cases of interest
here, the surface tension of the clouds overwhelms the gravitational
interaction with the stars and the stable final configuration would
have one star denuded and the other dressed with twice as many
neutrinos.  Binney and Tremaine~\cite{BINTRE} present the estimate that
2 stars, with $R\approx~R_\odot$, would actually collide once every
$10^{19}~\mbox{yr}.$  If, however, clouds extend to
$10^{17}~\mbox{cm}.$ the ratio of geometric cross sections is
$\approx~2\times~10^{12}$, so the encounter rate would be about 2 in
$10^7~\mbox{yr}.$, which is relatively fast on Galactic timescales.

Note, however, that the cloud stays with one star or the other.  The
evolutionary result of such collisions would be that neutrino clouds
would be found only with a fraction of the stars and that that fraction
would be smaller in more densely populated regions.  Furthermore, the
simple argument presented above takes no account of other neutrino
generations.  The possibility that the Sun has remained with an
attendant cloud remains viable.

\section{More Generations}

The preceeding discussion has been confined to the artificial case of
one generation for pedagogical purposes, but, as the last sections
illustrate, it is not reasonable to expect a good description of nature
without allowing for at least three generations of light neutrinos.
Necessary as this is, the unknown vacuum masses and couplings to the
scalar field provide a large number of free parameters.  The general
arguments setting allowed ranges will not change, but the actual limits
may depend on the number of participating neutrino species and the
detailed nature of the coupling of the scalar to those species,
consequently those limits may change by a few orders of magnitude.  On
the positive side, the existence of three mass eigenstates allows for
structures on different scales and can avoid the possibilities of
apparent inconsistencies described above.

To illustrate some of the complexity that will arise when all three
generations of neutrinos are considered, it is instructive to examine
the case of two generations.  In principle, there could be several
scalars coupling to the various generations with arbitrary strengths,
but that problem is too unconstrained to be instructive.  We shall
consider two simple examples, one in which the coupling is proportional
to the vacuum mass and one in which the coupling has the same strength
to both generations.

We first consider the case where the coupling of a neutrino to the
scalar field is proportional to the vacuum mass.  Label the two
neutrinos by $h$ (for heavy) and $l$ (for light).  Then, if $g$ denotes
the coupling constant for $h$,
\begin{equation}
g_l = \frac{m^{(0)}_l}{m^{(0)}_h} \times g
\end{equation}
This gives equations for the effective masses
\begin{equation}
m^{*}_h = m^{(0)}_h - g\phi
\end{equation} 
and
\begin{equation}
m^{*}_l = m^{(0)}_l - g_l\phi
\end{equation}
If we now define $y_h$ and $y_l$ as in section 2, we obtain
\begin{equation}
y_h = (1-\frac{g}{m^{(0)}_h} \phi)
\end{equation}
and
\begin{equation}
y_l = (1-\frac{g_l}{m^{(0)}_l}\phi) = (1-\frac{g}{m^{(0)}_h}\phi)
\end{equation}
or
\begin{equation}
y_l=y_h
\end{equation}
That is, there is only one function y for both species.  We then define
the quantities $x_h$, $x_l$, $e_h$ and $e_l$ in analogous fashion.
Also define the ratio
\begin{equation}
r =\frac{ m^{(0)}_l}{m^{(0)}_h}
\end{equation}
With these definitions, the differential equation for $y$ becomes
\begin{equation}
\frac{d^2y}{dz^2} + \frac{dy}{dz} = -1+y( 1+\frac{K_0}{2}F)
\end{equation}
where
\begin{equation}
F=e_{F_h} x_{F_h} -y^2\ln\frac{e_{F_h}+x_{F_h}}{y}  
+r^4\left[e_{F_l}x_{F_l}-y^2\ln\frac{e_{F_l}+x_{F_l}}{y}\right]
\end{equation}
The factor of $r^4$ comes from one power for scaling the energy and
three powers for scaling the Fermi momentum $x_{F_l}$.

The solution of the differential equation is carried out in much the
same manner as discussed in section 3, with $e_{F_h}$ and $e_{F_l}$
playing the roles of Lagrange multipliers fixing the total numbers of
heavy and light neutrinos respectively.

For this system, the density distributions are similar with the species
with the larger $e_F$ extending out to larger $z$.  Of course, the
factor $r^3$ in the density and $r^4$ in both the energy and the
driving term of the differential equation mean that the heavy neutrino
dominates the system.  For the special case where $e_{F_l}~=~e_{F_h}$,
the densities track exactly and the system is equivalent to one
generation with $K_0'~=(1~+~r^4)~K_0$.  The possibility of different
values for the $e_{F_i}$ allows for the composite distribution to fall
more slowly at the surface than it would for one generation and, if $r$
is not very different from 1, the gravitational force will also fall
more slowly.

An example of this is given in Figure 8, in which we have chosen the
ratio of light to heavy vacuum masses, $r$, to be $1/2$ and the
remaining parameters to give $K_0~=~20$ for the light neutrino.  We
have kept the number of heavy neutrinos, $N_h$, approximately constant
and varied the number of light neutrinos, $N_l$, from $0$ to a value
slightly in excess of $N_h$.  Since both neutrino densities are
involved in the driving term given in Eq (35), $e_{F_h}$ must be varied
as well as $e_{F_l}$, hence the approximate constancy of $N_h$.  For
each case  we have plotted the individual number densities as well as
the gravitational forces, calculated as in Figure 5, as a function of
the scaled radius.

There is a particularly interesting set of configurations in which
$e_{F_h}$ is less than the value of $y$ corresponding to the infinite
system of light neutrinos with a given $e_{F_l}$.  In these cases,
which can occur as the Universe expands, outside the cloud of heavy
neutrinos $y$ approaches this asymptotic value from below.  These
configurations correspond to the formation of clouds of heavy neutrinos
in the infinite sea of light neutrinos at an epoch in which the light
neutrinos are still very relativistic.

The case in which the scalar couples to all generations with the same
strength allows for more radical changes in the density distributions,
and we now examine that situation for two generations.  Here,
\begin{equation}
m^* = m^{(0)} - g\phi
\end{equation}
for both generations, i.e. the magnitude of the shift is the same  
for both.  Hence, if the shift is significant for the heavy neutrino it
may cause the effective mass of the light neutrino to change sign.
This is like the case of a massless fermion gaining mass from the
presence of a scalar condensate in field theory; the density of heavy
neutrinos serves as the source of an external scalar field for the
light neutrinos (and vice versa, although with little effect).  Care
must be taken in the assignment of quantum numbers, but the energy,
given by
\begin{displaymath}
E=(m^{*2} + k^2)^{1/2}
\end{displaymath}
is well behaved.  If $y$ denotes $m^*_h/m^{(0)}_h$, then
\begin{equation}
y_l = 1-(1-y)/r
\end{equation}
\begin{equation}
e_l = (y^2_l +x^2_l)^{1/2}
\end{equation}

The solution of the differential equation proceeds as before, with one
subsidiary condition.  Since $e_l~>~1$ implies that it is energetically
favorable for the light neutrino to be at infinity, light neutrinos
will be excluded from such a region and $ x_{F_l}$  is set to $0$
unless the condition
\begin{equation}
-e_{F_l} < y_l < e_{F_l}
\end{equation}
is met.  For $r$ sufficiently small, one may obtain a solution in which
the heavy neutrinos occupy a sphere with radius $a$; from $a$ to some
larger radius, $b$, there are no neutrinos and $\phi$ is a linear
combination of modified spherical Bessel functions of the third kind of
order $0$; then light neutrinos occupy a spherical shell from $b$ to
some large radius $c$. An example of such a distribution is plotted in
Figure 9, along with the gravitational force, as a function of the
scaled radius.  Here the vacuum mass ratio was $r~=~0.1$, $K_0~=~400$
for the heavy neutrino and $N_l~\approx~\frac{1}{2}~N_h$.

In principle, if the shell of light neutrinos occurs at a large enough
radius, such a distribution could radically affect the radial
dependence of the gravitational force produced by the neutrinos.  In
practice, that will require a very delicate matching in the solution of
the differential equation (troublesome for theorists, not for nature)
which requires the development of techniques like those reported is
section 2 for the treatment of very large clouds for one generation
only.

Since this scalar field may well be a contributor to, but not the only
contributor to, the neutrino mass matrix, the real condition is
probably more complicated than either of these examples.  The general
feature that the heaviest mass eigenstate should be most dense and of
smallest extent, common in both special scenarios, ought, however, to
persist.  If the dominant component of a neutrino cloud in the region
of one to several AU from the Sun were one of the heavier mass states,
say the medium mass eigenstate, with a small component to be the
interaction eigenstate associated with the electron, it could be
possible to provide enough energy density to affect satellite behavior
without violating the bounds from Supernova~1987a or conflicting with
Tritium beta decay experiments. In the latter case, while the effect
could extend for several tens of electron volts to either side of the
end point, a small mixing angle would provide only a small Pauli
blocking below and a concomitant small count rate above.

Whatever the detailed couplings might be, the effective mass of each of
the mass eigenstates will be affected by the distributions of all
three.  This could, therefore, alter the usual analyses of the MSW
effect in the sun and, since the neutrino densities associated with
supernovae is so very large, could lead to very different effects
there.  Recent analyses of the abundances produced in a neutron rich
exterior of supernovae~\cite{FUL} seems to favor an unusual spectrum of
masses, which might be achieved, for effective masses, by the scalar
field discussed here.

\section{Conclusions}

We have applied the techniques of Quantum Hadrodynamics to the study of
a system of neutrinos interacting through a light, weakly coupled
scalar boson.  We have shown that, for a wide range of parameters,
neutrinos will tend to condense into clouds, with dimensions the scale
of the  inverse boson mass.  In fact, for parameters which cause no
conflict with laboratory measurements, such clouds could easily be the
right size and density to affect experiments on and around the earth.

We have shown that it is likely that any such condensation would have
occurred before recombination and that the formation of neutrino clouds
could form a natural seeding mechanism for the formation of hadronic
objects on the scale of stars.  Neutrino cloud formation, being a phase
change, occurs very quickly, so these seeds are available at the
earliest possible epoch for star formation.

The extension of this work to more than one neutrino flavor depends on
the mass hierarchies of the neutrinos and the scalars as well as the
details of the coupling of each scalar to different generations.  For
the case of two generations of neutrinos and one scalar field, we have
looked at two simple choices for the couplings.  While different in
detail, both generate concentric spherical (assumed) distributions with
the lighter neutrinos (as determined by vacuum mass) extending farther
out.  We would expect this general feature to survive for three
generations, raising the possibility of the heaviest species being
essentially within the star, the other two occupying different regions
of space out to a distance, depending on the detailed history of the
system, of a fraction of a parsec.

If the density of the electron component of the neutrinos and
antineutrinos around the Sun is high enough, there could be observable
effects on very sensitive experiments such as the study of Tritium beta
decay to search for antineutrino mass effects or double beta decay
measurements seeking evidence that neutrinos are Majorana particles.

One consequence of the existence of such an interaction would be that
all such measurements would have to be interpreted in terms of
effective masses, rather than the vacuum masses that are relevant to
model building.

Whether terrestrial effects are observed or not, evidence for or
against the existence of such a scalar interaction is most likely to
come from astronomy and astrophysics.  The implications of the
existence of neutrino clouds, with respect to both the time scale and
the distribution of sizes, should be amenable to testing through
modelling and observation.  The gravitational effects within our own
Solar system, while subtle, could be observable in very high accuracy
satellite tracking data.  Depending on the precise model for several
generations, one may be able to observe the modifications of
oscillation and propagation in the extremely dense neutrino fluxes
associated with supernovae.

Whatever the experimental outcome of such tests may be, this problem
remains as a fascinating extension of the theoretical techniques of
QHD, developed for the study of atomic nuclei, to vastly different
regions of parameter space.

This work was supported in part by the US Department of Energy, the
National Science Foundation and the Australian ARC.  Some of the work
was carried out during visits to the Institute for Nuclear Theory at
the University of Washington; their hospitality is gratefully
acknowledged.  One of us (GJS) particularly wants to thank C.  Horowitz
for several very useful conversations.

 \newpage

\section*{Figure Captions}
{\bf Figure 1.} Scaled energies vs. scaled Fermi momentum for infinite
matter at zero temperature, $K_0 = 20$.  The energies and the Fermi
momentum are given in units of the neutrino vacuum mass, $\Mn $, $K_0$
is defined following Equation (7).  The dashed curve is the Fermi
energy and the solid curve is the average total energy per neutrino,
which includes the contribution of the static scalar field,
$e~=~e_{\nu}~+~e_s$.\\

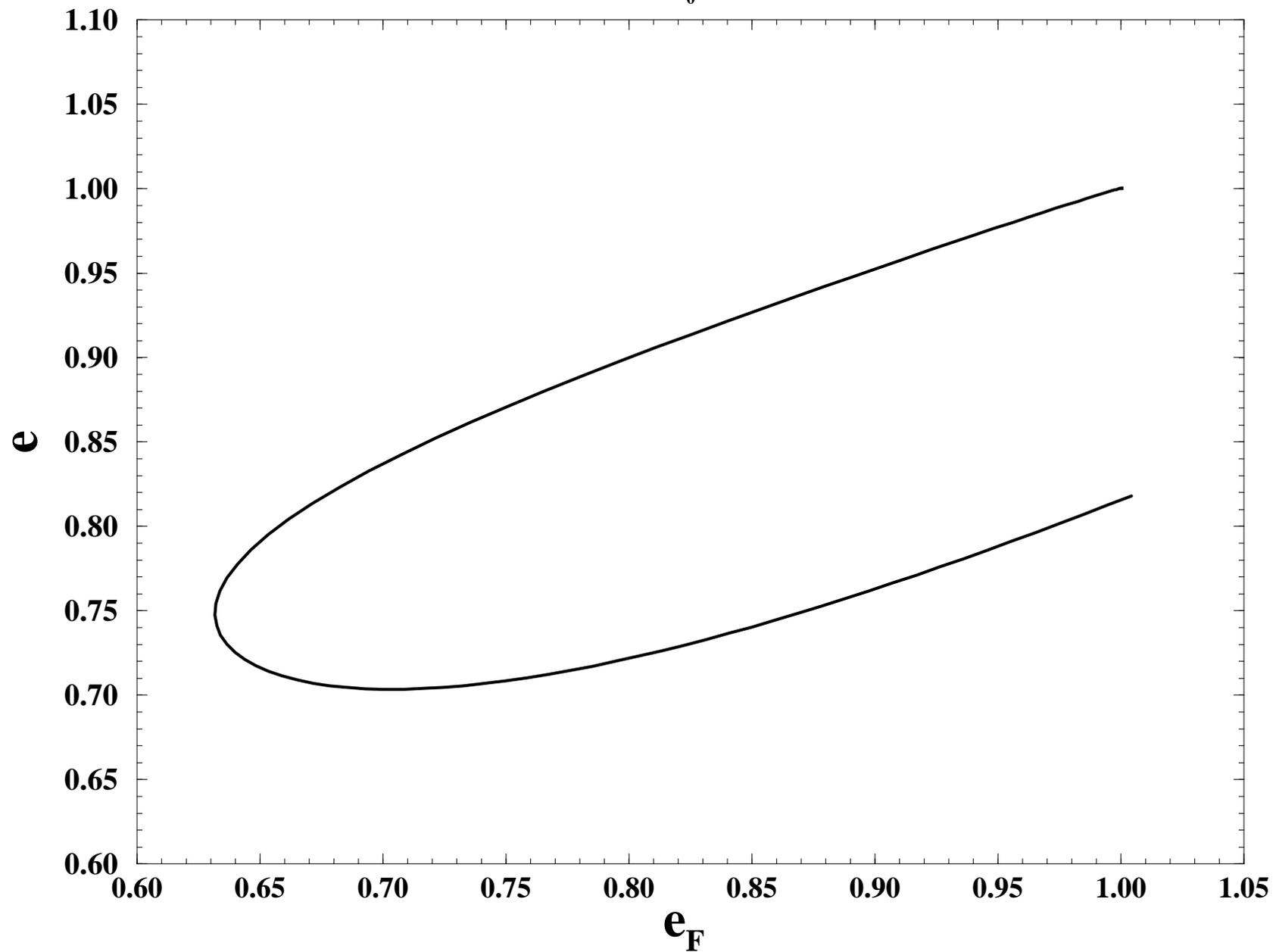
\noindent {\bf Figure 2.} Scaled total energy per neutrino vs. scaled
Fermi energy for infinite matter at zero temperature, $K_0~=~20$.\\

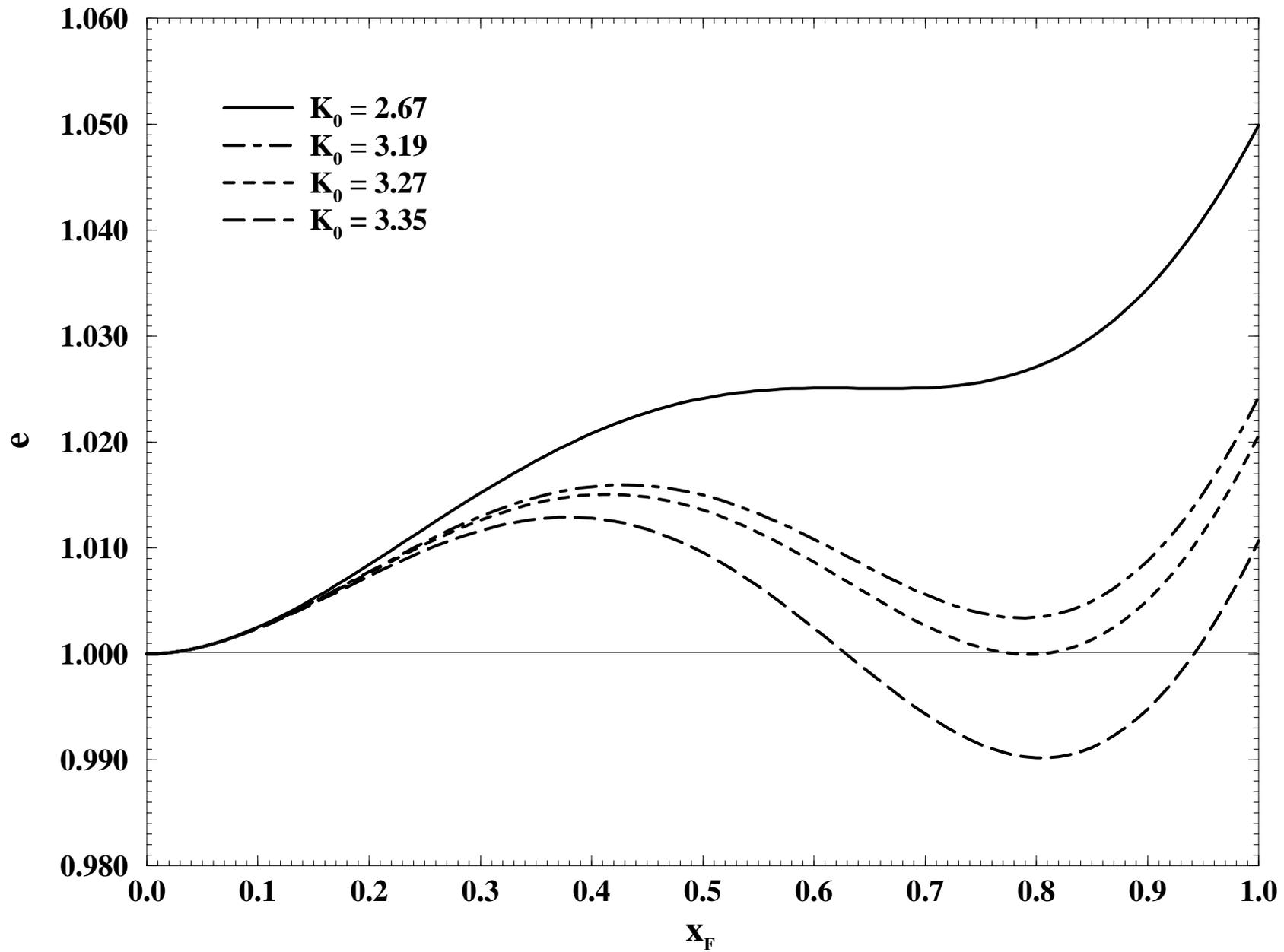
\noindent {\bf Figure 3.} Scaled total energy per neutrino vs. scaled
Fermi energy for infinite matter at zero temperature for several small
values of $K_0$.  For values of $K_0~>~3.27$, the system is
self-bound.  For $2.67~<~K_0~<~3.27$ there is a local minimum above
$e~=~1$, allowing for the possibility of metastable states.\\

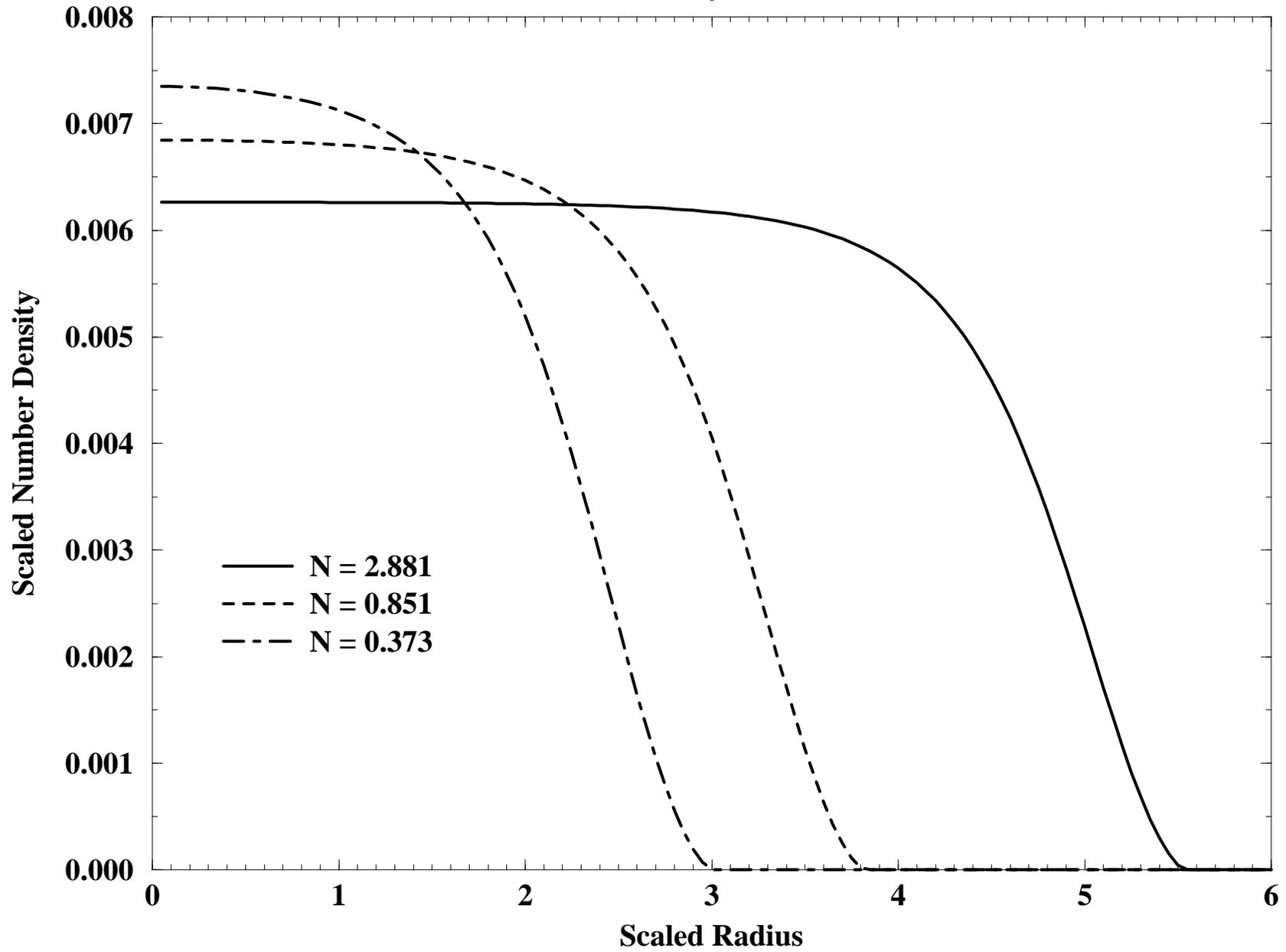
\noindent {\bf Figure 4.} Number density $\rho$ vs. scaled radius for
three different sized clouds for $K_0~=~20$.  In each case $e_F$ was
chosen, a solution was obtained as described in the text and the total
number N was obtained by integrating the density.  The radius is given
in units of $1/m_s$; the number in units of  $\mu^{-3}$.  For the solid
curve, $e_F~=~.74$ and $N~=~2.881$; for the dashed curve, $e_F~=~.76$
and $N~=~.851$; and for the dot-dashed curve, $e_F~=~.78$ and
$N~=~.373$.\\

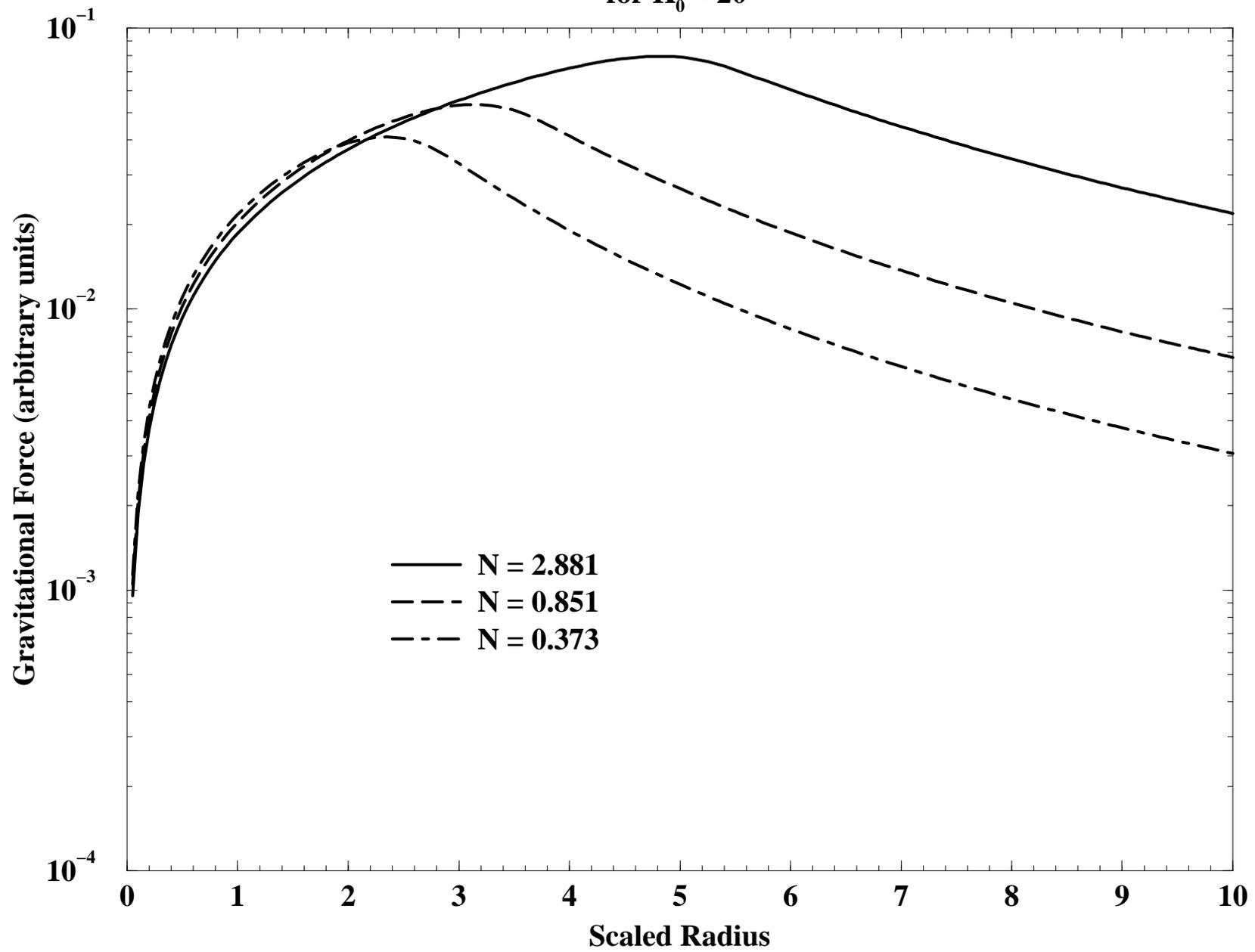
\noindent {\bf Figure 5.} Gravitational force, in arbitrary units, for
the three cases shown in Fig. 4.  $f_g$ at $z$ is given by
$\frac{1}{z^2}~\int_{0}^{z}~e~z^2~dz$.\\

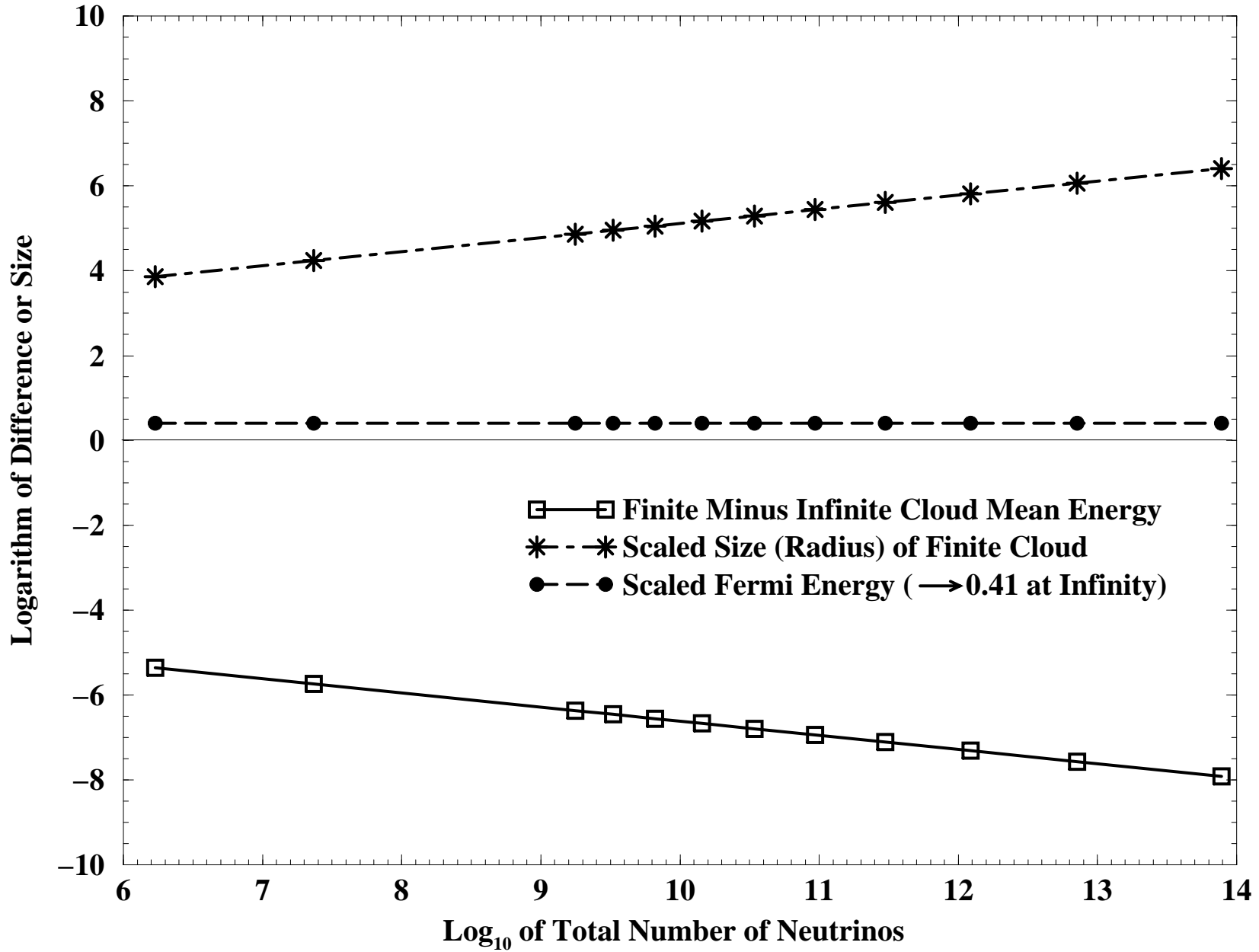
\noindent {\bf Figure 6.} Dependence of the average energy per neutrino
and of the radius on the number of neutrinos in the cloud, shown for
$K_0~=~200$, on a log-log plot.  The open squares are the values
obtained for the logarithm of the difference between $e$ for the cloud
and its value for infinite matter; the solid line has slope $-1/3$.
The stars are values of the logarithm of the radius at which the
neutrino density goes to zero; the dot-dash line has slope $1/3$.  The
filled circles are the values of the $e_F$ associated with each
calculation, the dotted curve is just a guide for the eye.\\

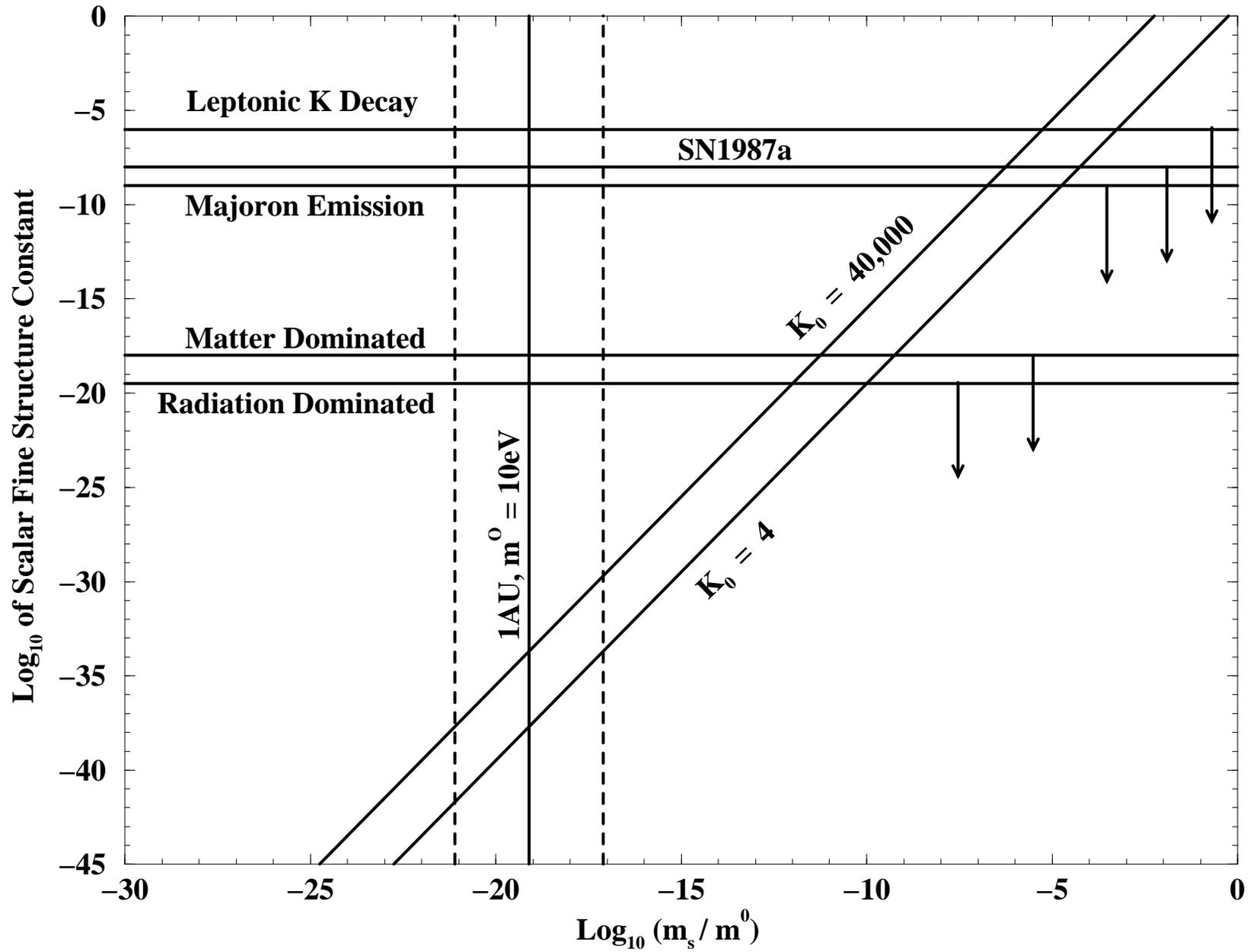
\noindent {\bf Figure 7.} Allowed regions of the $\mu$ -- $\tilde\alpha$
plane.  The horizontal lines indicate upper limits on $\tilde\alpha$
obtained from the labeled processes as discussed in the text.  The
solid vertical line corresponds to the value of $\mu$ for
$m_s~\simeq~1.3~\times~10^{-18}~eV$, appropriate to a range of $1AU$,
and $m^{0}_{\nu}~=~10 eV$.  The dashed vertical lines correspond to
changing $\mu$ by two orders a magnitude each way.  The diagonal lines
represent constant $K_0$, the lower for $K_0~=~4$ and the upper for
$K_0~=~40,000$.

\pagebreak

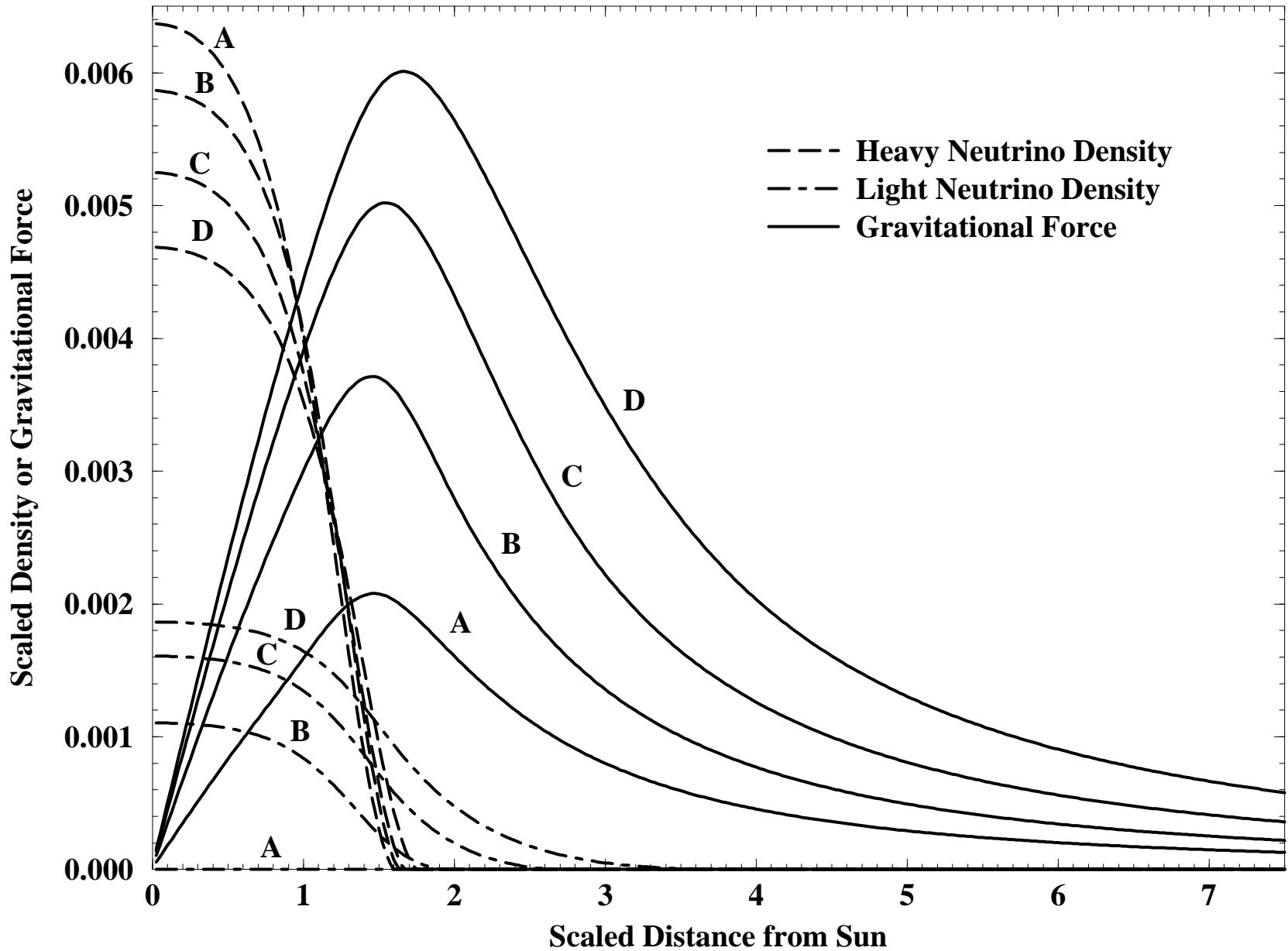
\noindent {\bf Figure 8.} Densities and gravitational force for two
neutrino mass eigenstates with $g~\propto~m$.  For this example,
$\frac{m^{0}_{l}}{m^{0}_{h}}~=~\frac {1}{2}$ and $K_0~=~20$ for the
light neutrino.  The four cases keep $N_h$ approximately constant and
vary $N_l$.  The coupling of  both densities  to the scalar field
requires $e_{F_h}$ to vary, hence the approximate equality.  Case A:
$N_h~=~.040$, $N_l~=~0$, and $e_{F_h}~=~.75$.  Case B:  $N_h~=~.042$,
$N_l~=~.011$, $e_{F_h}~=~.725$ and $e_{F_l}~=~.825$.  Case C:
$N_h~=~.040$, $N_l~=~.027$, $e_{F_h}~=~.7$ and $e_{F_l}~=~.93$.  Case
D: $N_h~=~.041$, $N_l~=~.051$,$e_{F_h}~=~.675$ and $e_{F_l}~=~.975$.
For each case, the long-dashed line gives the density for the heavy
neutrino, the dot-dashed line gives the density of the light neutrino
and the solid line gives the gravitational force due to the total
distribution calculated as for Fig. 5.  While the units of the
gravitational force are arbitrary, the relative strengths scale
properly.\\

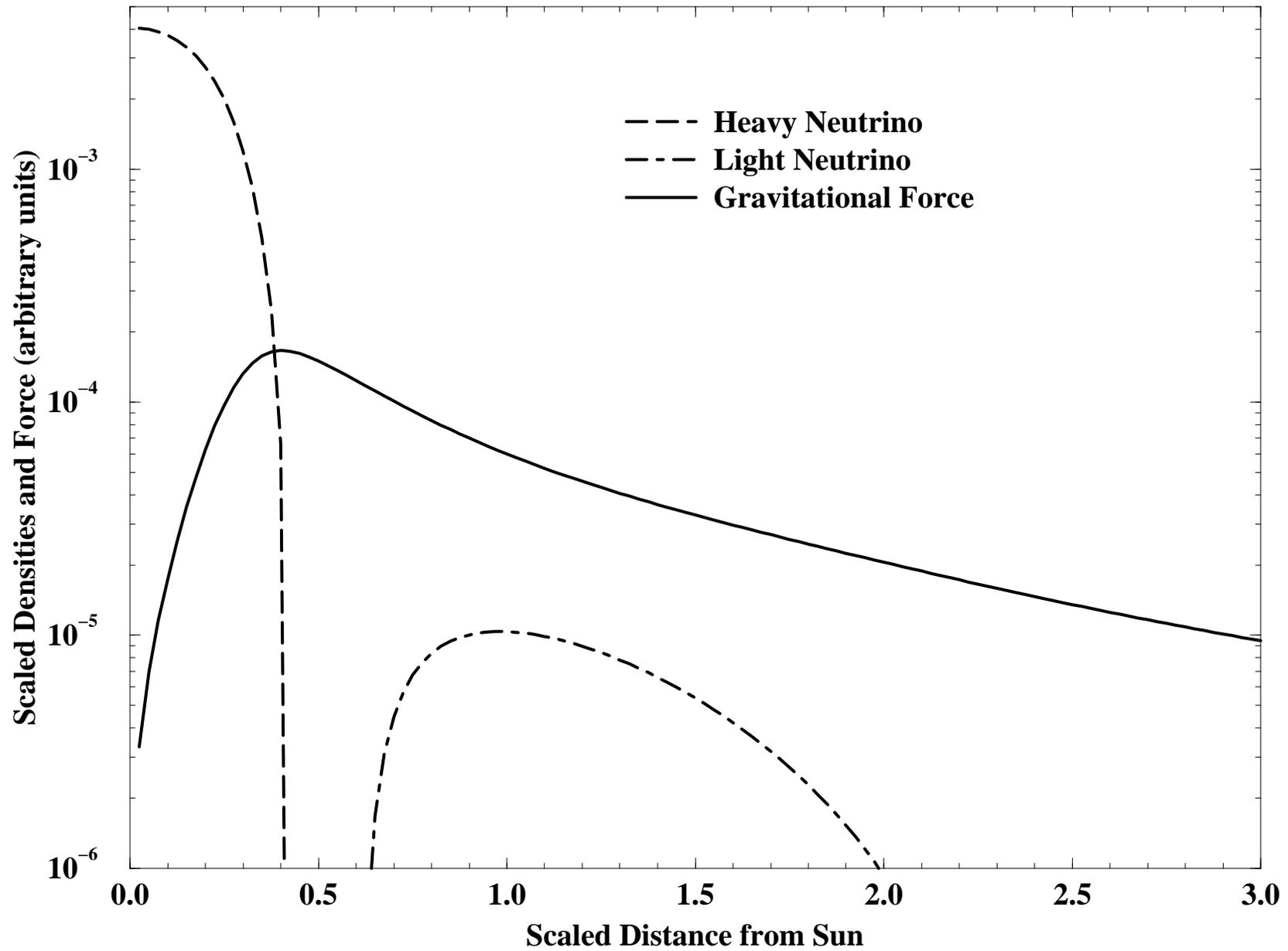
\noindent {\bf Figure 9.} Densities and gravitational force for two
neutrino mass eigenstates with constant coupling to the scalar field.
$K_0~=~400$ for the heavy neutrino and
$\frac{m^{0}_{h}}{m^{0}_{l}}~=~.1$  The curves have the same meaning as
in Fig. 8.  $e_{F_h}~=~.7$, $e_{F_l}~=~.85$, $N_h~=~.000339$ and
$N_l~=~.00162$.


\newpage

\begin{thebibliography}{99}

\bibitem{MKY} M. Kawasaki, H. Murayama and T. Yanagida, Mod. Phys.
Lett. \underline{A7}, 563 (1992).
\bibitem{TRE} R.A. Malaney, G.D. Starkman and S.  Tremaine,  Phys. Rev.
\underline{D51}, 324 (1995).  \bibitem{QHD} Brian D. Serot and John
Dirk Walecka, Adv.  in Nucl. Phys. Vol. \underline{16}, 1 (J.W. Negele
and Eric Vogt, eds. Plenum Press, NY 1986).
\bibitem{TRIT1} R. G. H. Robertson, T. J. Bowles, G. J.  Stephenson
Jr., D. L. Wark, J. F. Wilkerson and D. A. Knapp, Phys. Rev. Lett.
\underline {67}, 957 (1991).
\bibitem{TRIT2} E. Holzschuh, M. Fritschi and W. Kundig, Phys.  Lett.
\underline{B287}, 381 (1992).
\bibitem{TRIT3} Ch. Weinheimer, M. Prsyrembel, H. Backe, H.  Barth, J.
Bonn, B. Degen, Th. Edling, H. Fischer, L. Fleischmann, J. U. Grooss,
R. Haid, A. Hermanni, G. Kube, P leiderer, Th. Loeken, A. Molz, R. B.
Moore, A.  Osipowicz, E. W. Otten, A. Ricard, M. Schrader and M.
Steininger, Phys Lett.  \underline{B300}, 210 (1993).
\bibitem{TRIT4} H. Kawami, S. Kato, T. Ohshima, S. shibata, K Ukai, N.
Morikawa, N. Nogowa, K. Haga, T Nagafuchi, M. Shigeta, Y. Fukushima and
T. Taniguchi, Phys. Lett. \underline{B256}, 105 (1991).
\bibitem{TRIT5} Wolfgang Stoeffl and Daniel J. Decman, Phys.  Rev.
Lett.  \underline{75}, 3237 (1995).
\bibitem{TRIT6} A. I. Belesev, A. I. Bleule. E. V. Geraskin, A. A.
Golubev, O. V. Kazachenko, E. P. Kiev, Yu. E. Kuznetsov, V. M.
Lobashev, B. M. Ovchinnikov, V. I. Parfenov, I. V. Sekachev, A. P.
Solodukhin, N. A. Titov, I. E. Yarykin, Yu. I. Zakharov, S. N. Balashov
and P.  E. Spivak, INR preprint 862/94, Moscow (1994).
\bibitem{EA} E. Fischbach, G. T. Gillies, D. E. Krause, J. G.  Schwan
and C. Talmadge, Metrologia 29 (1992) 213 .  \bibitem{HZ}    E. R.
Harrison, Phys. Rev. D 1 (1970) 2726; Ya. B. Zel'dovich, Mon. Not. Roy.
Astron. Soc. 160 (1972) 1.
\bibitem{AA} C. Alcock {\em et. al.}, Nature (London), \underline{365},
621 (1993);  E. Aubourg {\em et. al.}, Nature(London), \underline{365},
623 (1993).
\bibitem{AFH} R.A. Alpher, J.W. Follin, Jr., and R.C.  Herman, Phys.
Rev. \underline{92},1347 (1953).  
\bibitem{SN87a} See references in {\em Review of Particle Properties},
Particle Data Group, Phys. Rev.  \underline{D50}, 1173 (1994).
\bibitem{GKS} T. Goldman, E.W. Kolb and G.J. Stephenson Jr., Phys Rev
\underline{D26}, 2503 (1981).
\bibitem{BLP} V. Barger, W. Y. Keung and S. Pakvasa, Phys. Rev.
\underline{D25}, 907 (1982).  
\bibitem{DBL1} Michael Moe and Petr Vogel, "Double Beta Decay" in {\it
Ann.  Rev. Nucl. Part. Sci.} \underline{44}, 247 (1994).
\bibitem{KTW} Edward W. Kolb, Michael S. Turner and Terrance P.Walker,
Phys. Rev. \underline{D34}, 2197, (1986).  \bibitem{HXS} See e. g., W.
C. Haxton and G.J. Stephenson Jr., "Double Beta Decay" in {\it Progress
in Particle and Nuclear Physics}, Vol.  \underline {12}, 409 (Sir Denys
Wilkinson, ed., Permagon Press, New York 1984).
\bibitem{MMN} Michael Martin Nieto, John D. Anderson, T.  Goldman,
Eunice L. Lau, and J.  P\'erez-Mercader, ``Theoretical Motivation for
Gravitation Experiments on Ultra-low Energy Antiprotons and
Antihydrogen'', in Proceedings of the {\it Third Biennial Conference on
Low-Energy Antiproton Physics, LEAP'94}, ed. by G. Kernel, P.  Krizan,
and M. Mikuz (World Scientific, Singapore, 1995), p. 606.
\bibitem{SWB} Steven Weinberg, Phys. Rev. \underline{128}, 1457 (1962);
Karl-Erik Berqvist, Nucl. Phys. \underline{B39}, 317 (1972).
\bibitem{BINTRE} James Binney and Scott Tremaine, {\em{Galactic
Dynamics}}, Princeton University Press, Princeton, NJ (1987).
\bibitem{FUL} G.M. Fuller, J. R. Primack and Y. -Z. Qian, Phys. Rev.
\underline{D52}, 1288 (1995).

\end{thebibliography}
\end{document}